\documentclass[a4paper,11pt]{article}

\usepackage[utf8]{inputenc}
\usepackage[T1]{fontenc}
\usepackage{graphicx}
\usepackage[textheight=243mm,textwidth=162mm,top=27mm,headheight=15pt,headsep=26pt,footskip=32pt]{geometry}
\usepackage{hyperref}

\title{\huge The most frequent programming mistakes\\
that cause software vulnerabilities}

\author{Raul Barbosa%
~~~~Frederico Cerveira%
~~~~Lu\'{i}s Gon\c{c}alo%
~~~~Henrique Madeira\\
CISUC, Department of Informatics Engineering\\
University of Coimbra\\
P-3030 290, Coimbra, Portugal}

\date{}

\begin{document}

\maketitle

\thispagestyle{empty}
\pagestyle{empty}

\begin{abstract}
All computer programs have flaws, some of which can be exploited to gain unauthorized access to computer systems. We conducted a field study on publicly reported vulnerabilities affecting three open source software projects in widespread use. This paper highlights the main observations and conclusions from the field data collected in the study.
\end{abstract}

\section{Introduction}

Programming mistakes sometimes create weaknesses that can be exploited to compromise information security. Such \textit{software security vulnerabilities} are the root of security risks in computer systems and this motivates the observation that \textit{``security problems are just bugs''} -- a well known quote by Linus Torvalds at the Linux Kernel Mailing List.

Perhaps contrary to intuition, programmers are not very creative in the ways they make mistakes. Simple mistakes, such as a missing \texttt{if} construct and statements, are by far the most frequent. As the saying goes, \textit{``great minds think alike and fools seldom differ''}. We examined 147 publicly available vulnerabilities in systems software (Linux, Xen and OpenSSH) and classified each one according to the Orthogonal Defect Classification method~\cite{Chillarege1996}.

\section{Frequent causes of vulnerabilities in C programs}

\begin{figure}[h]
\centering
\includegraphics[width=.95\textwidth]{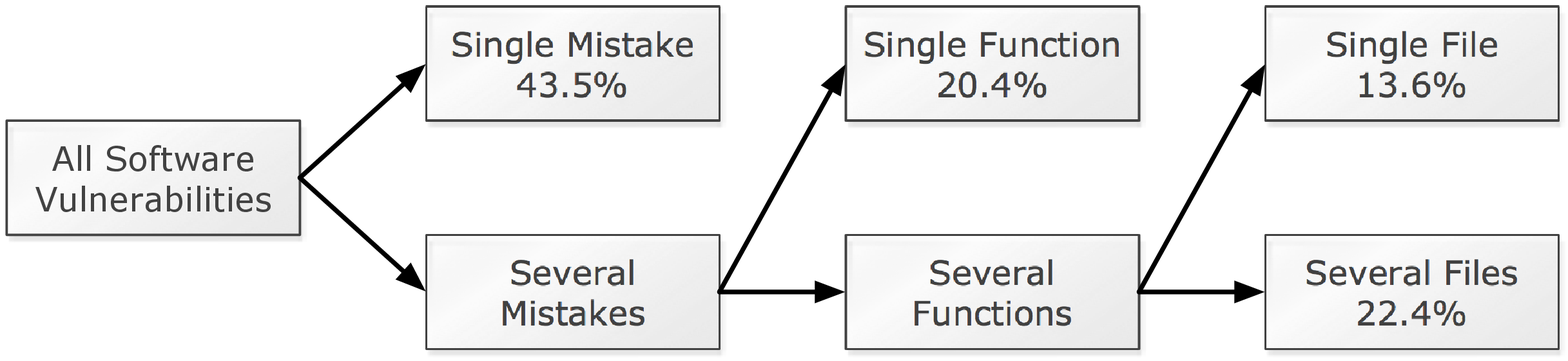}
\caption{Vulnerabilities range from single programming mistakes to defects involving several files.}
\label{fig:overview}
\end{figure}

The results of the \href{http://doi.org/10.1007/s00607-018-0657-y}{field study}~\cite{Barbosa2018} indicate that a typical security vulnerability involves a single function. Figure~\ref{fig:overview} depicts the diversity of software vulnerabilities, which are confined to a single function in 64\% of the cases (43.5\% + 20.4\%) but in the remaining cases involve several functions and possibly several different source code files.

Another relevant observation is that \textit{the majority of vulnerabilities consist of up to 3 programming mistakes}. In other words, simple unitary mistakes are sometimes combined to make up a security vulnerability. Around 75\% of vulnerabilities are composed of three or less mistakes.

The unitary programming mistakes made by programmers fall into few representative categories. Namely, \textit{missing \texttt{if} construct and statements}, \textit{missing function call} and \textit{extraneous function call} are present in the majority of software vulnerabilities. Moreover, \textit{erroneous expression in branch condition} is also a fairly common class of mistakes, with several variants. Table~\ref{tab:most_frequent} lists the most frequent programming mistakes found by examining real software vulnerabilities.

\begin{table}[!ht]
\centering
\caption{Most frequent programming mistakes occurring in software vulnerabilities.\label{tab:most_frequent}}
\begin{tabular}{|c|l|c|c|}
\hline
Mistake & \multicolumn{1}{c|}{Description}                           & Frequency \\ \hline\hline
MIFS & Missing \texttt{if} construct and statements                    & 38.8\%       \\ \hline
MFC  & Missing function call                                   & 21.8\%       \\ \hline
EFC  & Extraneous function call                                 & 10.9\%       \\ \hline
WLEC & Wrong logical expression used as branch condition          & 8.2\%       \\ \hline
EIFS & Extraneous \texttt{if} construct and statements                 & 6.8\%       \\ \hline
MLOC & Missing \texttt{|| Expr} in expression used as branch condition     & 6.8\%       \\ \hline
MLAC & Missing \texttt{\&\& Expr} in expression used as branch condition    & 6.8\%       \\ \hline
WALR & Wrong algorithm -- code was misplaced                   &  6.1\%       \\ \hline
MVAV & Missing variable assignment using a value             &  6.1\%       \\ \hline
\end{tabular}
\end{table}

\vspace{1mm}Table~\ref{tab:most_frequent} lists the nine most common mistakes that are at the root of security vulnerabilities of projects written in the C programming language. There are only a few key patterns. Missing \texttt{if} construct and statements is a mistake present in 38.8\% of the vulnerabilities examined. Missing or extraneous function calls are present in 21.8\% and 10.9\% respectively. Erroneous expression in branch condition also appears frequently, by merging all of the mistakes that involve some kind of error in a branch condition.

\section{Conclusion}

The field study summarized in this paper examined software security vulnerabilities in software written in the C programming language. Vulnerabilities published in CVE databases were examined. Specifically, each CVE entry contains a \textit{patch file} with the exact modifications needed to remove a vulnerability. Classifying those modifications according to the Orthogonal Defect Classification method allowed us to identify a few relevant patterns.

A security vulnerability is typically composed of up to three programming mistakes (this is the case in 75\% of the cases). In most cases, a vulnerability is confined to a single function (around 64\% of the cases) although in some cases multiple functions, possibly in different source code files, are involved. Lastly, most programming mistakes fall into few relevant patterns: missing \texttt{if} construct and statements; missing or extraneous function call; and erroneous expression used in branch condition. Consequently, the majority of software security vulnerabilities are caused by relatively simple mistakes, although a significant minority (about one out of four) involves uncommon mistakes or different functions spread over different source files. Advanced techniques to detect these will be necessary in the coming years to build secure software.

\bibliographystyle{apalike}
\bibliography{references}

\begin{thebibliography}{}

\bibitem[Barbosa et~al., 2018]{Barbosa2018}
Barbosa, R., Cerveira, F., Gon{\c{c}}alo, L., and Madeira, H. (2018).
\newblock Emulating representative software vulnerabilities using field data.
\newblock {\em Computing}.
\newblock Springer.

\bibitem[Chillarege, 1996]{Chillarege1996}
Chillarege, R. (1996).
\newblock Orthogonal defect classification.
\newblock In Lyu, M.~R., editor, {\em Handbook of Software Reliability
  Engineering}, chapter~9, pages 359--399. IEEE CS Press and McGraw-Hill.

\end{thebibliography}

\end{document}